\documentclass[11pt]{article}
\usepackage{epsfig}
\usepackage{amsmath}

\newlength{\dinwidth}
\newlength{\dinmargin}
\newcommand{\tdm}[1]{\mbox{\boldmath $#1$}}
\newcommand{\abar}{\bar\alpha_s}
\newcommand{\vx}{\tdm{x}}
\newcommand{\vy}{\tdm{y}}
\newcommand{\vz}{\tdm{z}}
\newcommand{\vk}{\tdm{k}}
\newcommand{\vr}{\tdm{r}}
\newcommand{\be}{\begin{equation}}
\newcommand{\ee}{\end{equation}}

\begin{document}
\begin{titlepage}
\begin{flushright}
\end{flushright}

\vspace*{1cm}

\begin{center}
{\LARGE \bf Nonlinear evolution of pomeron and odderon\\ 
in momentum space} \\
\vspace*{1cm}
\large Leszek Motyka\\[0.5cm]

{\it DESY Theory Group, Hamburg, Germany\\
\it and \\
\it Institute of Physics, Jagellonian University, Krak\'{o}w, Poland \\}
\vspace{1.5cm}

\end{center}

\begin{abstract}
The small~$x$ evolution of the QCD pomeron and the QCD odderon is investigated 
in the mean field limit of the Color Glass Condensate. 
The resulting system of coupled nonlinear evolution equations is 
transformed to the momentum space and analyzed at a very small momentum 
transfer. The main properties of the $C$-even and $C$-odd dipole densities 
in momentum space are obtained analytically. 
The critical scaling dimension is found for the odderon and the universal 
asymptotic behavior of the solutions is determined for small and large 
gluon momenta. We find that the same saturation scale 
characterizes both the pomeron and the odderon and both densities depend 
on the momentum only through the geometric scaling variable. The absorptive
effects are found to cause a strong suppression of the odderon exchange
amplitude for momenta below saturation scale and only a moderate suppression
for larger momenta.
\end{abstract}

\thispagestyle{empty}

\end{titlepage}

\section{Introduction}

A striking feature of strong scattering amplitudes at high energies 
is an overwhelming dominance of the exchange with the vacuum quantum numbers --
the pomeron. The exchange of the $C$-odd partner of the pomeron -- 
the odderon -- is much more elusive. 
In fact, despite experimental efforts only some weak
indications were found for the odderon contribution in hadronic 
processes (for a recent review see Ref.~\cite{ewerz}).
The extraordinary weakness of the odderon exchange is somewhat puzzling.
Part of the explanation of this fact is provided in a natural way by 
general color group arguments, since one needs at least two operators with the 
gluon quantum numbers to build the pomeron and at least three to build
the odderon. Therefore, the high power of $\alpha_s$ entering the odderon 
exchange amplitude suppresses it with respect to the $C$-even amplitude. 
Furthermore, the perturbative QCD pomeron amplitude grows steeply with energy, 
in contrast to a rather flat dependence of the perturbative odderon on the 
energy. Indeed, the existing theoretical estimates of the cross sections 
for various odderon mediated 
processes~\cite{etac,ryskin,ggeta,heid,brodsky,bbcv,twopi} predict small
cross sections, below the sensitivity of current experiments.  The only 
exception, for which some evidence of the odderon contribution was probably
measured, is the elastic $pp$ scattering at non-zero momentum transfer 
and at ISR energies~\cite{ppodd}.
Recently, it was realized that in addition 
to the mechanism described above, an important suppression of the odderon 
is caused by absorption of the odderon in a dense partonic 
system~\cite{ksw,hiim}.

In the high energy limit of perturbative QCD, the dynamics of the two gluon 
system (the pomeron) is described by the BFKL equation~\cite{bfkl}, 
which relies on a systematic resummation of perturbative corrections enhanced 
by powers of $\log (1/x)$, assuming $\alpha_s \ll 1$, 
and $\alpha_s \log (1/x) \sim 1$. 
The evolution of the pomeron amplitude at small~$x$ leads to a power like
growth of the amplitude, eventually giving rise to sizable unitarity 
corrections which tame the growth.
The perturbative small~$x$ evolution equation of the pomeron amplitudes, 
taking into account non-linear unitarity corrections  was derived by 
Balitsky~\cite{bal} and Kovchegov~\cite{kov} for a small and 
dilute projectile probing a dense and extended target. In the diagrammatic 
representation, the Balitsky-Kovchegov~(BK) equation resums BFKL pomeron fan 
diagrams in the large $N_c$ limit and it conforms with unitarity 
constraints at each impact parameter. The BK equation may be also obtained
as the mean-field limit of the effective theory of small~$x$ gluons in the
hadron wavefunction, that is the Color Glass Condensate approach~\cite{jimwalk}.

The perturbative realization of the odderon consists of at 
least three $t$-channel gluons in the color singlet state. 
The small~$x$ evolution equation of the odderon (the BKP equation) 
was derived long time ago~\cite{bartels,kp} but until recent years the 
properties of the solution have not been known. 
The spectrum of the QCD odderon Hamiltonian in 
the space of normalizable functions was found~\cite{jw} using the 
holomorphic symmetry of the problem. The odderon intercept appeared
to be smaller than unity, meaning that $C$-odd amplitudes should 
decrease with energy. In a following analysis, however, it was 
discovered~\cite{blv} that the normalizability condition of the holomorphic
wave functions may be relaxed, leading to a new odderon solution, having
a flat overall dependence on the collision energy. The new solution 
has a non-trivial property -- two reggeized $t$-channel gluons occupy the
same position in the transverse plane. Both the spectrum and
the eigenfunctions of this solution are contained by the standard
BFKL set of eigenfunctions with odd conformal spins.

The new odderon solution amplitude has a simple representation in the
dipole model framework~\cite{ksw}, where it is constructed using 
a Mueller dipole cascade model~\cite{mueller} with an initial condition with 
odd spatial parity. The QCD dipole cascade interacts with the target by 
an exchange of three elementary gluons in the $C$-odd color singlet state. 
In the dipole picture, the rescattering effects of the odderon amplitude were
obtained~\cite{ksw}, in analogy with the Kovchegov derivation of 
the~BK equation~\cite{kov}. 
The simultaneous investigation of the odderon and pomeron 
exchange amplitudes in the Color Glass Condensate lead to a generalization 
for the abovementioned results in the form of coupled non-linear equations 
involving both the $C$-even and $C$-odd amplitudes~\cite{hiim}. 
In this letter we shall study in detail the 
impact of the absorption on the momentum distribution of the gluons 
in the odderon.


\section{Formalism}

Effects of absorption of the odderon in high energy scattering were considered 
by Kovchegov, Szymanowski, Wallon~\cite{ksw} in the framework of dipole model.
They found a correction term to the odderon evolution equation 
bilinear in the pomeron and odderon densities. Hatta, Iancu, Itakura 
and McLerran~\cite{hiim} analyzed $C$-even and $C$-odd
scattering amplitudes using the formalism of the Color Glass 
Condensate. The energy evolution of the amplitudes was described by 
a system of functional equations. In the mean field limit 
the odderon evolution equation agreed with the result of Ref.~\cite{ksw}. 
A new non-linear term was found, however, contributing to the pomeron 
evolution beyond the BK~equation, quadratic in the odderon amplitude. 
Eventually, in the leading logarithmic approximation the small~$x$ 
evolution equations of the $C$-even amplitude $N(\vx,\vy;\tau)$ and the 
$C$-odd amplitude $O(\vx,\vy;\tau)$ in the transverse position plane 
read~\cite{hiim}:
\[
{\partial N(\vx,\vy;\tau) \over \partial \tau} = 
{\abar \over 2\pi} \int d^2 z \;
{ (\vx - \vy)^2 \over (\vx - \vz)^2  (\vz - \vy)^2} 
\left[
N(\vx,\vz;\tau) + N(\vz,\vy;\tau) - N(\vx,\vy;\tau) \right.
\]
\be
\left.
- N(\vx,\vz;\tau)N(\vz,\vy;\tau)  + 
O (\vx,\vz;\tau) O(\vz,\vy;\tau) 
\right],
\label{pom}
\ee
\[
{\partial O(\vx,\vy;\tau) \over \partial \tau} = 
{\abar \over 2\pi} \int d^2 z \; 
{ (\vx - \vy)^2 \over (\vx - \vz)^2  (\vz - \vy)^2} 
\left[ 
O(\vx,\vz;\tau) + O(\vz,\vy;\tau) - O(\vx,\vy;\tau) 
\right.
\]
\be
\left.
- O(\vx,\vz;\tau)N(\vz,\vy;\tau)  -
N (\vx,\vz;\tau) O(\vz,\vy;\tau) 
\right],
\label{odd}
\ee
with $\tau = \log (1/x)$ and $\vx$, $\vy$ and $\vz$ representing positions of 
the end points of color dipoles in the transverse plane. Let us refer to
this system as the WHIMIKS equation, after the authors' initials.

\begin{figure}
\begin{center}
\epsfig{file=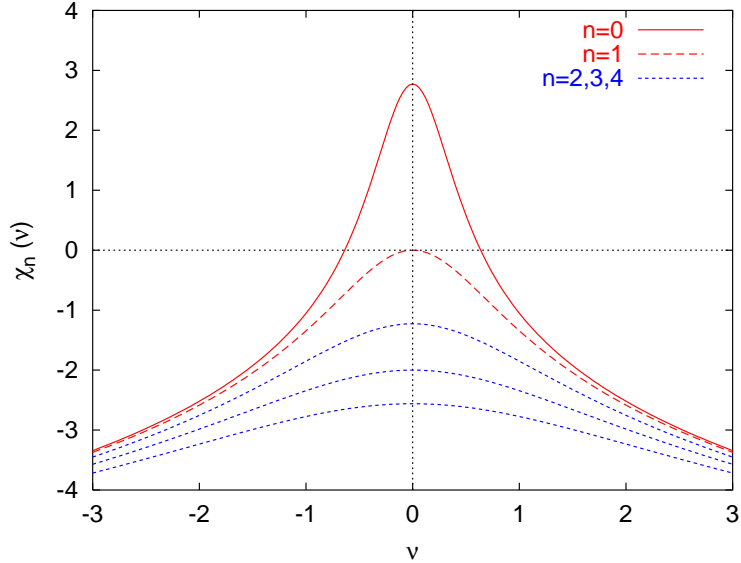, width=0.65\columnwidth } 
\end{center}
\caption{\em Eigenvalues of the BFKL kernel $\chi_n (1/2 + i\nu)$
for conformal spins enumerated by $n$. The values of 
$\bar\alpha_s\chi_n(1/2+i\nu)$ for $\nu=0$ with $n=0$ and $n=1$
give the intercept of the pomeron and the odderon correspondingly.}
\end{figure}

By construction, the pomeron and the odderon exchange amplitudes have definite 
parities with respect to exchange of the gluon positions, that is:
\be
N(\vy,\vx;\tau) =  N(\vx,\vy;\tau), \qquad O(\vy,\vx;\tau) =  -O(\vx,\vy;\tau).
\label{parity}
\ee
For the sake of simplicity, we restrict our analysis of the nonlinear evolution 
equations to the translationally invariant case, corresponding to the limit 
of the target size being much larger than the dipole size. 
In this limit, the impact parameter dependence of 
$N(\vx,\vy;\tau)$ and $O(\vy,\vx;\tau)$ 
may be neglected. 
Thus, we assume
\be
N(\vx,\vy;\tau) = N(\vy-\vx,\tau), \qquad O(\vx,\vy;\tau) = O(\vy-\vx,\tau).
\ee

In the next part, we choose to work in the momentum space. 
Therefore we define the momentum dependent functions $\Phi(\vk,\tau)$ and 
$\Psi(\vk,\tau)$ 
describing the pomeron and the odderon exchange respectively
\be
\Phi(\vk,\tau) = \int {d^2 \vr \over 2 \pi r^2} N(\vr,\tau) \exp(-i \vk\vr), 
\ee
\be
\Psi(\vk,\tau) = \int {d^2 \vr \over 2 \pi r^2} O(\vr,\tau) \exp(-i \vk\vr). 
\ee
In the dipole formalism $\Phi(\vk,\tau)$ and $\Psi(\vk,\tau)$ represent dipole densities
in momentum space. It is straightforward to obtain the form of equations 
(\ref{pom}) and (\ref{odd}) in this representation, 
\be
{\partial \Phi(\vk,\tau) \over \partial \tau} = 
\abar\, (K'\otimes \Phi)(\vk,\tau) -  \abar \Phi^2(\vk,\tau) + \abar \Psi^2(\vk,\tau),
\label{kpom}
\ee
\be
{\partial \Psi(\vk,\tau) \over \partial \tau} = \abar\, (K'\otimes \Psi)(\vk,\tau) -  
2 \abar \Phi(\vk,\tau)\Psi(\vk,\tau),
\label{kodd}
\ee
with the linear kernel being related to the standard LL BFKL kernel. To be more
specific, the action of the kernel $K'$ on the function $f(k^2)$ is given 
by the BFKL kernel action on the function $k^2\,f(k^2)$, for instance
\be
(K'\otimes \Phi)(\vk,\tau) = \int { d^2 \vk' \over (\vk - \vk')^2} \, 
\left[ 
\Phi(\vk',\tau) - {k^2\Phi(\vk,\tau) \over 
\vk'^2 + (\vk - \vk')^2} 
\right].
\ee
In the near-forward scattering limit (the momentum transfer $q$ being much 
smaller than typical gluon momenta) the eigenfunctions $f_{n,\gamma}(\vk)$ 
of the kernel $K'$ take the simple form in the polar coordinates $(k,\varphi)$,
\be
f_{n,\gamma} (k,\varphi) = f_{n,\gamma} k^{2\gamma} \cos(n\varphi) 
\label{eigenf}
\ee
and the eigenvalues of the BFKL kernel read
\be
\chi_n(\gamma) = 2\psi(1) - \psi(|n|/2 + \gamma) - \psi(|n|/2 + 1 - \gamma).
\label{chin}
\ee
with $n$ and $\gamma$ being the conformal spin and the scaling dimension 
respectively. Therefore, the solution to the linear equation 
\be
{\partial f(k,\varphi;\tau) \over \partial \tau} = 
\bar\alpha_s [K' \otimes f](k,\varphi;\tau)
\ee
may be expressed as
\be
f(k,\varphi;\tau) = \sum_{n=0}^{\infty} \int_{-1/2-i\infty}^{-1/2+i\infty} 
{d\gamma \over 2\pi i}\; f^{(0)} _{n,\gamma}\; k^{2\gamma} \cos(n\varphi)\,
e^{\bar\alpha_s\chi_n(\gamma+1)\tau}
\label{bfsol}
\ee
where the coefficients  $f^{(0)} _{n,\gamma}$ are obtained by a projection
of the initial condition $f^{(0)}(k,\varphi) = f(k,\varphi;\tau=0)$ on the
kernel eigenfunctions (\ref{eigenf}).
The rapidity dependence of a solution with a given conformal spin $n$ at large
rapidities is driven by the behavior of the corresponding kernel eigenvalue 
in the vicinity of the point $\gamma=-1/2$ where the saddle
point of the integration resides (note the shifted argument
of $\chi_n$ in~(\ref{bfsol})). 
The dependence of $\chi_n(\gamma)$ along
the integration contour is illustrated in Fig.~1 for $n=0,1,2,3$ and $n=4$,
as a function of $\nu$, where $\gamma=1/2+i\nu$. 
For conformal spins $n \geq 2$ the eigenvalue $\chi_n(1/2+i\nu) < 0$ for all 
$\nu$ (see Fig.~1) thus the components with conformal spins $n \geq 2$ 
are exponentially suppressed with increasing rapidity.
The leading component in the pomeron sector, with $n=0$, 
grows exponentially with rapidity, with the famous BFKL intercept 
$\omega_0 = \chi_0(1/2) = 4 \log 2\, \abar$. 
Due to the parity property~(\ref{parity}) 
in the transverse space, the even conformal spins contribute exclusively to 
the pomeron, and the odd conformal spins to the odderon. 
The dominant component in the odderon sector at large rapidities corresponds 
to $n=1$, with the intercept $\omega_1 = \chi_1(1/2) = 0$ and the main effect 
of the evolution of the component is a diffusion in the transverse momentum. 

Taking into account the rapidity dependence of the components with conformal 
spins $n\geq 2$ it is justified to retain only the leading conformal spins,
that is 
\be
\Phi(\vk,\tau) = \Phi(k,\tau), \qquad \Psi(\vk,\tau) = 
\Psi(k,\tau)\, \cos(\varphi),
\ee
with $\varphi$ being the angle between $\vk$ and a characteristic direction in the 
transverse plane, given by the $C$-odd impact factor or by a small but non-vanishing 
momentum transfer. 

Evolution equations (\ref{pom}) and (\ref{odd}) are consistent with this Ansatz,
except of the quadratic odderon term in (\ref{pom}). Thus, we approximate the
contribution of this term by its projection on the $n=0$ (constant)  
angular component $\Psi^2 (k)\cos^2 (\varphi) \to (1/2)\, \Psi^2 (k)$.  
In this approximation we obtain the following system of equations 
\be
{\partial \Phi(k,\tau) \over \partial \tau} = 
\abar\, 
\int_0 ^\infty {dk'^2 \over k'^2} \left[
{k'^2 \Phi(k',\tau) - k^2 \Phi(k,\tau) \over |k'^2-k^2|}
+ { k^2 \Phi(k,\tau) \over \sqrt{4k'^2 + k^2}} 
\right]
-\abar \Phi^2(k,\tau) + 
 {1\over 2}\abar \Psi^2(k,\tau), 
\label{eqphi}
\ee
\be
{\partial \Psi(k,\tau) \over \partial \tau} = 
\abar\, 
\int_0 ^\infty {dk'^2 \over k'^2} \left[
{ k'^2\,(k_< / k_>)\,  \Psi(k',\tau) - k^2 \Psi(k,\tau) \over |k'^2-k^2|}
+ { k^2 \Psi(k,\tau) \over \sqrt{4k'^2 + k^2}}
\right]
-2 \abar \Phi(k,\tau)\Psi(k,\tau),
\label{eqpsi}
\ee
where $k_< = \min (k,k')$ and $k_> = \max (k,k')$. 
These equations form the basis of an ongoing numerical 
analysis~\cite{lmprepare}.


\section{Analytical properties of the solution}

Let us consider the following Ansatz for the solutions of equations
(\ref{eqphi},\ref{eqpsi}), in analogy to the similar Ansatz applied in 
the case of the Balitsky-Kovchegov equation~\cite{glr,barlev,gms}
\be
\Phi(k,\tau) = \int {d\gamma \over 2\pi i}\, e^{\gamma t} \; e^{\omega_P (\gamma) \tau}
\; \phi(\gamma), 
\label{imelpom}
\ee
\be
\Psi(k,\tau) =  \int {d\gamma \over 2\pi i}\, e^{\gamma t} \; e^{\omega_O (\gamma) \tau}
\; \psi(\gamma),
\label{imelodd}
\ee
with $ t=\log(k^2/k_0^2)$.
and $\omega_P (\gamma)$, $\omega_O (\gamma)$ and $\phi(\gamma)$,  
$\psi(\gamma)$ being the unknown functions, for which the system is 
solved.  Functions $\omega_R (\gamma)$ with $R=P,O$ determine the 
rapidity dependence of the component with the scaling dimension $\gamma$.

From the Mellin transform of (\ref{eqphi},\ref{eqpsi}) we get
\be
\omega_P(\gamma) \phi(\gamma) e^{\omega_P(\gamma)\tau}  
= \abar \chi_0(\gamma+1) \phi(\gamma) e^{\omega_P(\gamma)\tau} -
\abar \int { d\gamma' \over 2\pi i} \, \phi(\gamma-\gamma')\, \phi(\gamma') 
e^{[\omega_P(\gamma') + \omega_P(\gamma-\gamma')]\tau},  
\label{omp}
\ee
and
\be
\omega_O(\gamma) \psi(\gamma) e^{\omega_O(\gamma)\tau}  
= \abar \chi_1(\gamma+1)\psi(\gamma) e^{\omega_O(\gamma)\tau}  -
2\abar \int { d\gamma' \over 2\pi i}\, \phi(\gamma-\gamma')\, \psi(\gamma') 
e^{[\omega_O(\gamma') + \omega_P(\gamma-\gamma')]\tau}, 
\label{omo}
\ee
where we omitted the subdominant nonlinearity due to $\Psi^2(k,\tau)$ 
in (\ref{omp}). When the amplitudes are small, the nonlinearities 
may be neglected, leading to the following form of functions $\omega_{P,O}$ 
given by the linear (BFKL) part:
\be
\omega_P (\gamma) = \abar\chi_0(\gamma+1), \qquad 
\omega_O (\gamma) = \abar\chi_1(\gamma+1).
\label{omlin}
\ee

In the absence of the nonlinear rescattering terms, the behavior of 
equations (\ref{eqphi},\ref{eqpsi}) is well known. 
There, the solution at large rapidities 
$\tau \gg 1$ is dominated by the saddle point contribution of the inverse
Mellin transform, e.g.\
\be
\Phi(k,\tau) = \int_{-1/2-i\infty} ^{-1/2+i\infty} 
{d\gamma \over 2\pi i}\;
\phi_0(\gamma) 
\exp[\gamma \log(k^2/k_0^2)] \, 
\exp[\abar \chi_0(\gamma+1) \tau], 
\ee
with $\phi_0(\gamma)$ specified by the initial condition and $k_0$ being an 
arbitrary scale.
With $\tau \gg \log(k^2/k_0^2)$, the saddle point occurs at $\gamma=-1/2$
and 
\be
\Phi(k,\tau) \simeq {\phi_0(-1/2)\over  \sqrt{2\pi D_0 \tau}} \;
{k_0 \over k}\, \exp (\omega_0 \tau)\, 
\exp\biggl( - {\log^2(k^2/k_0^2)\over 2 D_0\tau} \biggr),
\label{aspom}
\ee
with the diffusion coefficient 
\be
D_0 = 28\zeta(3)\abar \simeq 33.66\abar.
\ee
Analogously, for the odderon solution
\be
\Psi(k,\tau) \simeq {\psi_0(-1/2)\over  \sqrt{2\pi D_1 \tau}} \;
{k_0 \over k}\, 
\exp \biggl( - {\log^2(k^2/k_0^2) \over 2D_1\tau} \biggr),
\label{asodd}
\ee
with 
\be
D_1 = 4\zeta(3)\abar \simeq 4.81 \abar \ll D_0.
\label{d1}
\ee
Note that the odderon solution has no exponential rapidity dependence,
as $\omega_1 = \abar\chi_1(1/2) = 0$, see (\ref{chin}). Interestingly
enough, the diffusion coefficient of the odderon solution is smaller
by a significant factor of seven than the diffusion coefficient of the pomeron.
Consequently, the saddle point approximation 
for the odderon (\ref{asodd}) is  valid for much larger rapidities $\tau$ 
than the asymptotic formula (\ref{aspom}) for the pomeron. 

Approximate solutions (\ref{aspom}) and (\ref{asodd}), valid at large $\tau$,
are driven by the Gaussian diffusion factor in $\log (k^2/k_0^2)$ distorted 
by the prefactor $1/k$. Thus, for any starting conditions  
$\Phi(k,\tau_0)$ and $\Psi(k,\tau_0)$, for sufficiently large $\tau$ the 
diffusion will populate with color dipoles the domain of small momenta, 
and both $\Phi(k,\tau)$ and $\Psi(k,\tau)$ will become large, switching on 
non-linear corrections for small $k$. 
Note, however, that the much slower diffusion of the hard odderon 
amplitude makes it safer against the infra-red effects with respect to the 
pomeron amplitude. The evolution length needed for the odderon to diffuse 
from the hard initial condition to the infra-red domain is more than~2.5 
times longer that it is for the pomeron. It also means that the scattering 
amplitude of a hard $C$-odd source should be less affected by absorptive
corrections than the $C$-even amplitude.  

\begin{figure}
\begin{center}
\epsfig{file=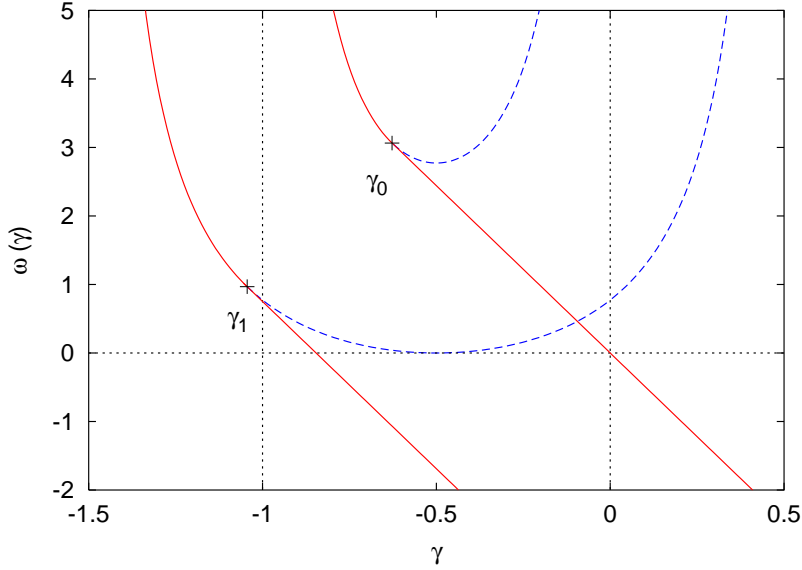, width=0.65\columnwidth } 
\end{center}
\caption{\em Matching of $\omega(\gamma)$ in the linear and saturated
regimes for the pomeron (upper set) and the odderon (lower set) dipole densities as functions of the real part of scaling dimension $\gamma$. The critical scaling dimensions are indicated for the pomeron ($\gamma_0$) and for the 
odderon ($\gamma_1$).  }
\end{figure}

In the saturation regime, the saddle point evaluation of the integrand 
over $\gamma'$ in the r.h.s of (\ref{omp}) leads to the following 
condition~\cite{glr,barlev,gms}
\be
\omega_P (\gamma) = 2  \omega_P (\gamma / 2 ),
\ee
with the solution
\be
\omega_P (\gamma) = C \gamma. 
\label{omp2}
\ee
Recall that such dependence of $\omega_P (\gamma)$ implies the geometric 
scaling property~\cite{glr,barlev,gs,gms},
\be
\Phi(k,\tau) = \int {d\gamma \over 2\pi i}\, \phi(\gamma)\, 
\exp(\gamma t + C\gamma \tau) = \Phi(k^2 \exp(C\tau)).
\ee

At sufficiently large rapidity $\tau$, the integral over $\gamma'$ in the r.h.s of 
(\ref{omo}) is also dominated by a contribution from the saddle 
point $\gamma'=\gamma_s(\gamma)$, such that 
\be
{\partial \over \partial \gamma'} 
\left. [\omega_O(\gamma') + \omega_P(\gamma-\gamma')] 
\right|_{\gamma'=\gamma_s(\gamma)} = 0.
\label{oddsaddle} 
\ee 
Then we obtain
\be
\omega_O(\gamma) = \omega_O(\gamma_s(\gamma)) + \omega_P(\gamma-\gamma_s(\gamma)),
\ee
thus, using (\ref{omp2})
\be
\omega_O(\gamma) - \omega_O(\gamma_s(\gamma)) =  C (\gamma - \gamma_s(\gamma)).
\ee
The equation is fulfilled for any $\gamma_s(\gamma)$ if $\omega_O(\gamma)$ is a 
linear function,
\be
\omega_O(\gamma)  =  A + C \gamma. 
\label{omo2}
\ee
After applying the inverse Mellin transform (\ref{imelodd}), one finds that
in the saturation regime 
\be
\Psi(k,\tau) =  \int {d\gamma \over 2\pi i}\, e^{\gamma t} \; 
e^{(A + C \gamma)\tau}\; \psi(\gamma) = \Psi(k^2\, \exp(C\tau))\, e^{A\tau},
\label{oddscal}
\ee
so $\Psi(k,\tau)$ depends on the transverse momentum only through the 
scaling variable $\xi = k^2\, \exp(C\tau)$ of the BK solution,  as
$\omega_P(\gamma)$ and  $\omega_O(\gamma)$ in the
saturation region are characterized by the same coefficient $C$.
In contrast to the pomeron case, however, it follows from (\ref{oddscal}) 
that the odderon dipole density $\Psi$ depends on rapidity not only 
through $\xi$ but also by an overall decreasing 
(as it will be shown that $A < 0$) exponential suppression factor.

Following the Refs.~\cite{barlev,gms} we impose the condition that of 
the smooth transition (with the first derivative) of 
$\omega_P(\gamma)$ between the linear and saturation regimes
given by (\ref{omlin}) and (\ref{omp2}) respectively. Then, the
transition point $\gamma_0$ is determined by    
\be
(\gamma_0 + 1) \, \chi_0 (\gamma_0 + 1) = \chi'_0 (\gamma_0 + 1),
\ee
thus $\gamma_0 \simeq -0.6275$ and 
$C = \abar \chi'_0 (1+\gamma_0) = -4.8834\abar$.
The analogous condition imposed on $\omega_O(\gamma)$ defines the 
transition point $\gamma_1$ between the linear and saturated regime in
the odderon sector. From (\ref{omp2}) and (\ref{omo2}) we deduce that
\be
\chi'_1(\gamma_1+1) = \chi'_0(\gamma_0+1). 
\ee
and $\gamma_1 \simeq - 1.0441$.
Now, it is straightforward to determine the coefficient $A$ in (\ref{omo2}):
\be
A = \abar [ \chi_1(\gamma_1 + 1) - (\gamma_1+1) \chi_1 '(\gamma_1+1)].   
\ee
The matching procedure is illustrated in Fig.~2.
The above results for $\omega_{P,O}(\gamma)$ in the saturated regime 
may be summarized in the following way,
\be
\omega_P(\gamma) = \abar\chi_0'(\gamma_0) \gamma \simeq 
-4.8834\, \abar \gamma,
\label{ompf}
\ee
\be
\omega_O(\gamma) = 
\abar[ \chi _1(\gamma_1+1) +  (\gamma - \gamma_1) \chi_1'(\gamma_1+1)] 
\simeq \abar [ -4.8834\, \gamma -4.1311].
\label{omof}
\ee

It was shown~\cite{kov} that at low $k$ the solution to the Balitsky-Kovchegov 
equation is dominated by $\gamma \simeq 0$, close to the pole of 
$\chi_0 (\gamma+1)$,
giving rise to the asymptotic behavior $\Phi(k,\tau) \simeq \log(Q_s(\tau)/k)$, 
with $Q_s(\tau)$ being the saturation scale.  
Analogously, the $C$-odd dipole density  $\Psi(k,\tau)$ in the low~$k$ 
region is dominated by the pole of $\chi_1 (\gamma+1)$ at $\gamma = 1/2$. 
This corresponds to the behavior $\Psi(k,\tau) \sim k/Q_s(\tau)$ at $k\to 0$. 
Indeed, for $\Phi(k,\tau) \sim \log k$ and  $\Psi(k,\tau) \sim k^{\alpha}$ 
the nonlinear term $\Phi(k,\tau) \Psi(k,\tau) \sim  k^{\alpha} \log k$ has
a double pole in the Mellin space $\sim 1/(\gamma-\alpha)^2$. 
In equation~(\ref{kodd}) this double pole can be  matched only by a contribution from
$(K' \otimes \Psi)(k,\tau)$, which in the Mellin space
has the pole structure of $\chi_1(\gamma+1)\psi(\gamma)$, see also (\ref{omo}). 
Thus, the position of the pole in $\psi(\gamma)$ must coincide with the anti-collinear 
pole of $\chi_1(\gamma+1)$ at $\gamma =1/2$.

At the large momentum asymptotics $k\gg Q_s(\tau)$, the nonlinearity in the 
evolution equation is not relevant, as both $\Phi(k,\tau)$ and  $\Psi(k,\tau)$ 
exhibit a power like decrease at large $k$. Therefore the large~$k$ tails
of  $\Psi(k,\tau)$ and  $\Phi(k,\tau)$ should be driven by the largest 
scaling dimensions $\gamma$ for which the linear term dominates the evolution
in equations (\ref{omp}) and (\ref{omo}). It follows that the 
large~$k$ behavior of $\Phi(k,\tau)$ and  $\Psi(k,\tau)$ is characterized by
the scaling dimensions corresponding to the matching points 
$\gamma_0$ and $\gamma_1$  respectively:
$\Phi(k,\tau) \sim k^{2\gamma_0}$ and  $\Psi(k,\tau)  \sim k^{2\gamma_1}$.
Note, however, that the asymptotical dependencies at a given value of~$k$ 
(small or large) are reached only after the evolution length $\tau$ is 
sufficiently long for the diffusion to occur from the initial condition 
to the region of momenta close to $k$.

Finally, let us determine the rapidity dependencies of the dipole densities at 
a fixed momentum $k$ and a large $\tau$. It follows from the previous considerations, 
that for $k < Q_s (\tau)$, $\Phi(k,\tau) \sim \log(\tau) + {\mathrm const}$ 
and the $C$-odd density is strongly suppressed,
$\Psi(k,\tau) \sim \exp((C/2+A)\tau) \simeq  \exp(-6.55\bar\alpha_s\tau)$.
At large $k$,  $\Phi(k,\tau)$ grows as 
$\exp(C\gamma_0\tau) \sim \exp (3.05\bar\alpha_s\tau)$ and  
$\Psi(k,\tau)$ decreases with rapidity rather mildly, 
$\Psi(k,\tau) \sim \exp((C\gamma_1+A)\tau) \simeq  \exp(-0.97\bar\alpha_s\tau)$. 

In the forthcoming study~\cite{lmprepare} the above results will be compared
to the results of numerical investigations.   


\section{Summary and remarks}

Let us recapitulate the approximate results:
\begin{enumerate}

\item 
The scale parameter (saturation scale) $Q_s(\tau)$ of the solutions to 
the nonlinear, coupled evolution equations for the pomeron and the 
odderon exchange depends exponentially on the rapidity 
$Q_s(\tau) \simeq Q_0\, \exp(|C|\tau/2)$, with the coefficient $C$  
that coincides with its analogue in the solution of the Balitsky-Kovchegov 
equation. 

\item 
The pomeron solution exhibits approximate geometric scaling
$\Phi(k,\tau) \simeq \Phi(\xi)$ with $\xi =  k^2\exp(C\tau)$, 
and the shape of the odderon solution depends on the same scaling variable 
$\xi$ but the overall normalization decreases with rapidity:
$\Psi(k,\tau) \simeq \Psi(\xi) \exp(A\tau)$.

\item In the saturation region $k\ll Q_s(\tau)$ one gets the following 
leading behavior of the solutions:
\[
\Phi(k,\tau) \simeq \log(Q_s(\tau)/k),\qquad\qquad 
\Psi(k,\tau) \sim k/Q_s(\tau)\; \exp(A\tau), 
\]
note that the normalizing prefactor of $\Psi(k,\tau)$ is not uniquely 
determined in our approach.

\item In the region of linear evolution $k\gg Q_s(\tau)$, one obtains   
\[
\Phi(k,\tau) \sim [k/Q_s(\tau)]^{2\gamma_0}, \qquad\qquad
\Psi(k,\tau) \sim [k/Q_s(\tau)]^{2\gamma_1}\,\exp(A\tau).
\]

\item The approximate numerical values of the parameters read:
$C \simeq -4.88\abar$, $A \simeq -4.11\abar$, $\gamma_0 \simeq -0.63$ and
$\gamma_1 \simeq -1.04$.

\end{enumerate}

Clearly, the accuracy of the applied approximations is limited, so the
obtained characteristics of the solution to the system of equations
(\ref{kpom}) and (\ref{kodd}) may differ from the exact results. For instance,
in the case of the BK equation, the value of the exponent 
$C \simeq -4.2\bar\alpha_s$ was determined in a numerical analysis, 
smaller than the value $C \simeq -4.88\bar\alpha_s$ found in the 
approximate scheme~\cite{gms}. The accuracy level of the present analysis is 
expected to be similar. Moreover, the studied evolution equations are given
in the leading $\log(1/x)$ approximation and thus the energy dependence of
the saturation scale is by far too steep. Consequently the absorptive
effects are exaggerated.

Let us also point out the importance of the squared odderon term
in the evolution equation (\ref{kpom}). This term may change qualitatively
the behavior of the solution for both the pomeron and the odderon in the 
case when the initial condition contains only the $C$-odd part, 
that is $\Phi(k,\tau=0)=0$. Then, in the absence of $\Psi^2(k,\tau)$,
the pomeron solution would vanish for all rapidities, $\Phi(k,\tau)=0$, and
the system (\ref{kpom},\ref{kodd}) would reduce to the linear odderon 
evolution equation. The quadratic linear term, however, acts as a source for 
the pomeron amplitude, driving it away from zero. Actually, due to positive 
rapidity dependence of $\Phi(k,\tau)$ and the negative one of $\Phi(k,\tau)$, 
at large rapidities the pomeron amplitude will dominate anyway, generating the
saturation scale and leading to the generic behavior of the both 
amplitudes, which was summarized above.

The results of our analysis may be compared to the approximate asymptotic
solution to the WHIMIKS equation obtained in Ref.~\cite{hiim},
\be
O(\vx, \vy) \simeq 
\exp\left[-{|C|\over 2} \bar\alpha_s^2 (\tau-\tau_0)^2\right],
\qquad \mathrm{for} \quad |\vx -\vy| \gg 1/Q_s(\tau).
\label{hiimlt}
\ee
We confirm the conclusion
of  Ref.~\cite{hiim} that the saturation scale generated in the pomeron
sector drives the absorptive effects in the odderon sector, imposing
its rapidity dependence on the $C$-odd scattering amplitude. 
It is clear, however, that the current analysis in the momentum space provides 
novel information about the universal features of the solution to the 
WHIMIKS equation, like the universal scaling behavior for 
$k\ll Q_s(\tau)$ and $k\gg Q_s(\tau)$ and the overall rapidity dependence.
Besides that, it should be useful to have the direct insight into the 
momentum dependence of the odderon exchange.

It should be mentioned that a similar approach to the one proposed in this 
letter may be applied to study the evolution of components with higher 
conformal spins of the BK equation or of the WHIMIKS equation. 
We anticipate that the nonlinear effects will cause strong damping of 
the higher conformal spin components, in analogy to the odderon $n=1$ component. 
Finally, it would be worthwhile to continue the analysis of WHIMIKS equation
keeping the non-zero momentum transfer and investigate the phenomenological
consequences of the absorption for some classical odderon mediated processes,
like, for instance, the $\eta_c$ photoproduction off a proton~\cite{etac,bbcv}.


\section{Conclusions}

The system of coupled nonlinear small~$x$ evolution equations for the pomeron 
and the odderon (WHIMIKS equation) was analyzed with approximate analytical
methods in momentum space. The system generates the 
saturation scale, growing exponentially with rapidity, with the same exponent 
as emerges from  the BK equation. With respect to the BK case, the change was 
not significant for properties of the $C$-even dipole density at large
rapidity. 
The $C$-odd dipole density was found to depend on the momentum only through
the geometric scaling variable of the BK equation, but in addition an overall
dependence on rapidity was found in the form of a decreasing exponential 
prefactor. Low and large momentum asymptotics of the dipole densities 
was determined. The main conclusion from this paper is that the effects of 
rescattering lead to a strong damping of the odderon for momenta below  
the saturation scale and a rather moderate suppression at large momenta.
Therefore, the absorptive effects reduce substantially the odderon 
contribution to semihard processes with a scale smaller than the saturation
scale, which may explain the observed weakness of the odderon exchange
amplitude.

\section*{Acknowledgments}
I would like to thank Jochen Bartels, Krzysztof Golec-Biernat and 
Augustin Sabio Vera for useful discussions and comments. The support
of a grant of the Polish State Committee for Scientific Research 
No.\ 1~P03B~028~28 is gratefully acknowledged.

\end{document}